\documentclass[prb,twocolumn,amsmath,amssymb, superscriptaddress]{revtex4}
\usepackage{graphicx}% Include figure files
\usepackage{dcolumn}% Align table columns on decimal point
\usepackage{bm}% bold math

\begin{document}

\preprint{APS/123-QED}

\title{Glasslike vs. crystalline thermal conductivity in carrier-tuned Ba$_8$Ga$_{16}$X$_{30}$ clathrates (X = Ge, Sn)}% Force line breaks with \\

\author{M. A. Avila}
% \email{avila@sci.hiroshima-u.ac.jp}
\author{K. Suekuni}%
\affiliation{%
Department of Quantum Matter, ADSM, Hiroshima University,
Higashi-Hiroshima 739-8530, Japan
}%
\author{K. Umeo}%
\affiliation{%
N-BARD, Hiroshima University, Higashi-Hiroshima 739-8526, Japan
}%
\author{H. Fukuoka}%
\author{S. Yamanaka}%
\affiliation{%
Department of Applied Chemistry, Graduate School of Engineering,
Hiroshima University, Higashi-Hiroshima 739-8527, Japan
}%
\author{T. Takabatake}%
\affiliation{%
Department of Quantum Matter, ADSM, Hiroshima University,
Higashi-Hiroshima 739-8530, Japan
}%
\affiliation{%
Institute for Advanced Materials Research, Hiroshima University,
Higashi-Hiroshima 739-8530, Japan
}%

\date{\today}% It is always \today, today,

\begin{abstract}
The present controversy over the origin of glasslike thermal
conductivity observed in certain crystalline materials is
addressed by studies on single-crystal x-ray diffraction, thermal
conductivity $\kappa(T)$ and specific heat $C_p(T)$ of
carrier-tuned Ba$_8$Ga$_{16}$X$_{30}$ (X = Ge, Sn) clathrates.
These crystals show radically different low-temperature
$\kappa(T)$ behaviors depending on whether their charge carriers
are electrons or holes, displaying the usual crystalline peak in
the former case and an anomalous glasslike plateau in the latter.
In contrast, $C_p(T)$ above 4 K and the general structural
properties are essentially insensitive to carrier tuning. We
analyze these combined results within the framework of a
Tunneling/Resonant/Rayleigh scatterings model, and conclude that
the evolution from crystalline to glasslike $\kappa(T)$ is
accompanied by an increase both in the effective density of
tunnelling states and in the resonant scattering level, while
neither one of these contributions can solely account for the
observed changes in the full temperature range. This suggests that
the most relevant factor which determines crystalline or glasslike
behavior is the coupling strength between the guest vibrational
modes and the frameworks with different charge carriers.

\end{abstract}

% PACS, the Physics and Astronomy Classification Scheme.


\maketitle

\section{Introduction}

The understanding of thermal conductivity behavior $\kappa(T)$ is
of direct interest to any research involving the discovery, design
and development of materials for thermoelectric conversion
applications, where $\kappa(T)$ should be as small as possible,
while at the same time thermopower $S(T)$ and electrical
conductivity $\sigma(T)$ should be large. In the semiclassical
theory for electron and phonon transport,\cite{ziman60a,tritt04a}
several mechanisms are known as contributors to heat
conduction/phonon scattering in a material, consequently affecting
its overall thermal conductivity.

In metals, heat conduction by charge carriers is the largest
contribution, and is well described by the Wiedemann-Franz law
$\kappa_c=L\sigma T$, which directly relates the carrier thermal
conductivity $\kappa_c$ with an appropriate Lorentz number $L\sim
2-3\times 10^{-8}$~W$\Omega$/K$^2$, the electrical conductivity
$\sigma$ and the temperature $T$. Due to their typically large
charge carrier densities $n_c$, metals have large $\sigma(n_c,T)$
and thus large $\kappa_c$ in the range of $50-500$~W/m~K at room
temperature. % It is also immediately clear that in the
% electron-phonon scattering regime, where metals have a roughly
% $T$-linear resistivity $\rho(T)=1/\sigma(T)$, they will show
% roughly temperature-independent $\kappa_c(T)$, which is indeed
% observed.

Conversely, semi-metallic, semi-conducting and insulating
compounds have low $n_c$ and $\sigma(T)$, therefore small and
often negligible $\kappa_c(T)$ and the overall heat conduction
behavior $\kappa(T)$ near room temperature is in the range of
$10-50$~W/m~K, governed mostly by contributions $\kappa_L(T)$
arising from the crystal lattice. At $T\rightarrow0$,
$\kappa(T)\rightarrow0$ from basic thermodynamic principles, so
within the first few Kelvins $\kappa(T)$ increases quickly as a
power law $T^r$, with $1\le r\le3$ depending on which phonon
scattering mechanisms dominate at low temperatures. At higher
temperatures the phonon scattering is generally described as
governed mostly by umklapp processes, for which the Debye
approximation approach shows a decrease with a $T^{-1}$
dependence. Therefore, at some intermediate temperature usually
around $10-50$~K, a characteristic ``crystalline peak'' is
observed in $\kappa(T)$ for common compounds, due to the crossover
from one regime to another.\cite{tritt04a}

The peak usually appears equally in polycrystalline materials
since grain boundary scattering is in general a minor
contribution,\cite{tritt04a,nol98a} unless the average grain size
becomes very small or the temperature very low. However, glasses
are an exception to the above because of two basic factors: the
very low mean free path for both electrons and phonons, and the
presence of low-energy tunnelling states (TS), i.e., different
localized potential minima for atomic positions in their amorphous
distribution of nuclei.\cite{tritt04a} This class of materials
shows extremely low heat conduction and a universal behavior of
$\kappa(T)$: the lowest temperature behavior (up to $\sim1$~K)
rises as $T^2$ due to scattering by the tunnelling states, then
levels off as a characteristic intermediate temperature plateau
(attributed to Rayleigh scattering). Above $\sim20$~K it resumes a
slow increase, until roughly levelling off again at higher
temperatures.

\begin{figure}[tb]
\includegraphics[angle=0,width=86mm]{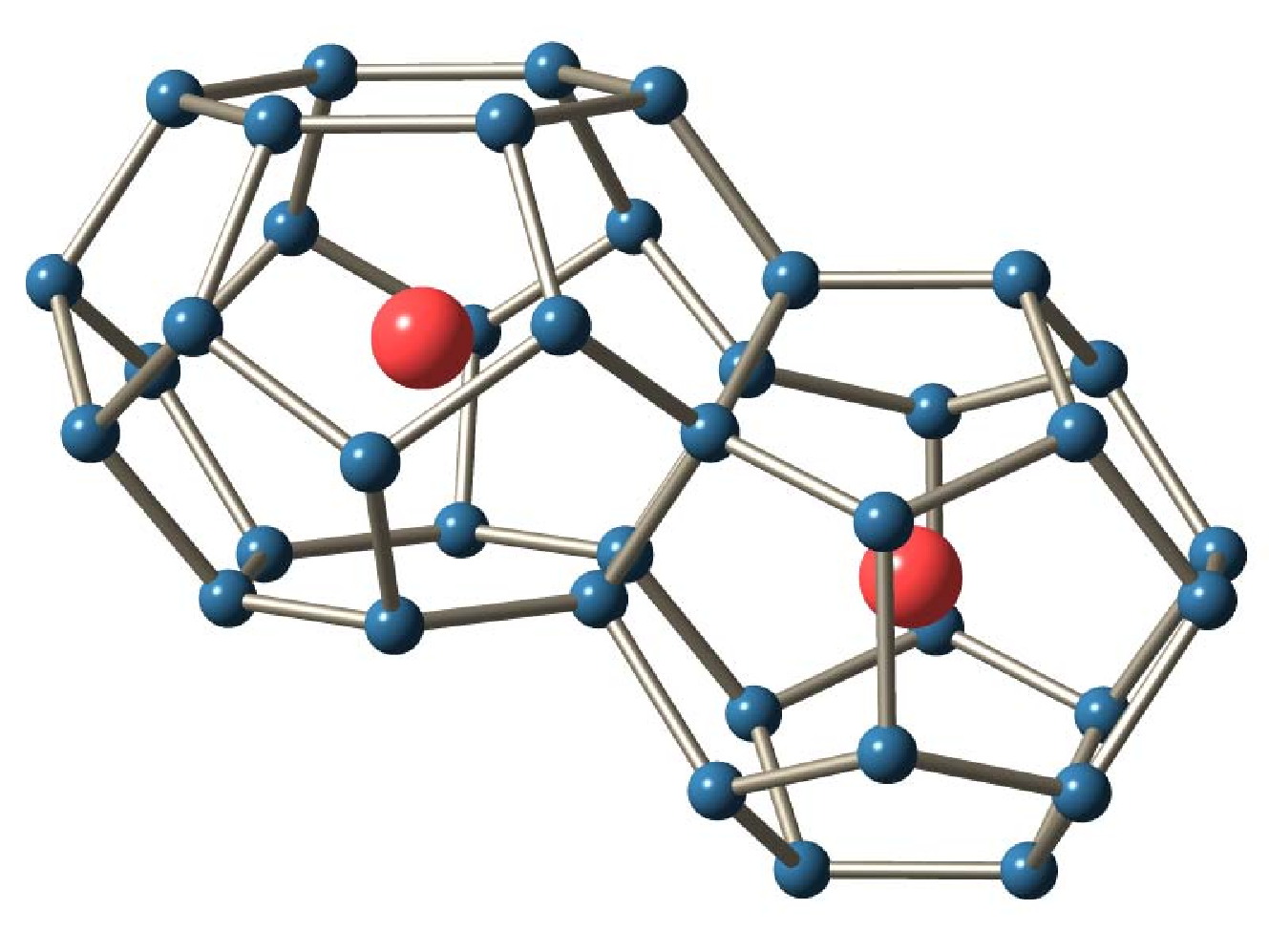}% Here is how to import EPS art
\caption{\label{T1Cages} (Color online) The two cages of the
type-I clathrate structure adopted by Ba$_8$Ga$_{16}$Ge$_{30}$. If
we consider the cage ``construction unit'' as 4 atoms connected in
zig-zag from top to bottom, then the larger $X_{24}$ cage (left)
is made of 6 such units and the smaller $X_{20}$ cage (right) is
made of 5 units.}
\end{figure}

In more recent years, the search for new and potentially useful
thermoelectric materials\cite{wood88a,sla95a} has led to the
discovery of compounds that not only have unusually low thermal
conductivity, but whose general behavior resembles that of a
glassy material despite the fact that they are true (albeit
disordered) crystalline lattices.\cite{cah92a} A prominent example
is the intermetallic compound Sr$_8$Ga$_{16}$Ge$_{30}$, with a
filled type-I clathrate structure\cite{eis86a} (6 larger $X_{24}$
cages forming tetrakaidecahedra plus 2 smaller $X_{20}$ cages
forming dodecahedra, see Fig.~\ref{T1Cages}) for which $\kappa(T)$
was first measured by Nolas \emph{et al.} in 1998.\cite{nol98a} A
model was proposed\cite{cohn99a} to explain this material's
glasslike behavior, based on the idea that TS exist for the Sr(2)
guest ion in the $X_{24}$ cage,\cite{nol00a} to which it is rather
loosely bound because of an ion-to-cage size mismatch. A
combination of phonon scattering by TS, resonant scattering on
large, Einstein-like localized vibration modes (guest
 \emph{rattling}) and Rayleigh scattering was used to adequately
reproduce the experimental $\kappa(T)$ behavior (henceforth we
will refer to this combination as the TRR model). Later
investigations amply demonstrated a splitting of the Sr(2) site
into four off-center positions,\cite{cha00a,sal01a,zer04a} among
which the ions could indeed tunnel.

As other clathrate compounds started being investigated, the TRR
model was challenged by at least two other models. One proposes
that the tunnelling states are not required, only an
\emph{off-center} vibration of the guest
ions,\cite{bri04a,baum05a} and another proposes that the guest
ions don't play a major role at all at low temperatures, but
rather it is the phonon scattering on charge carriers that leads
to the glasslike behavior.\cite{ben04a,ben05a,pach05a,ben06a}

\begin{figure}[tb]
\includegraphics[angle=0,width=60mm]{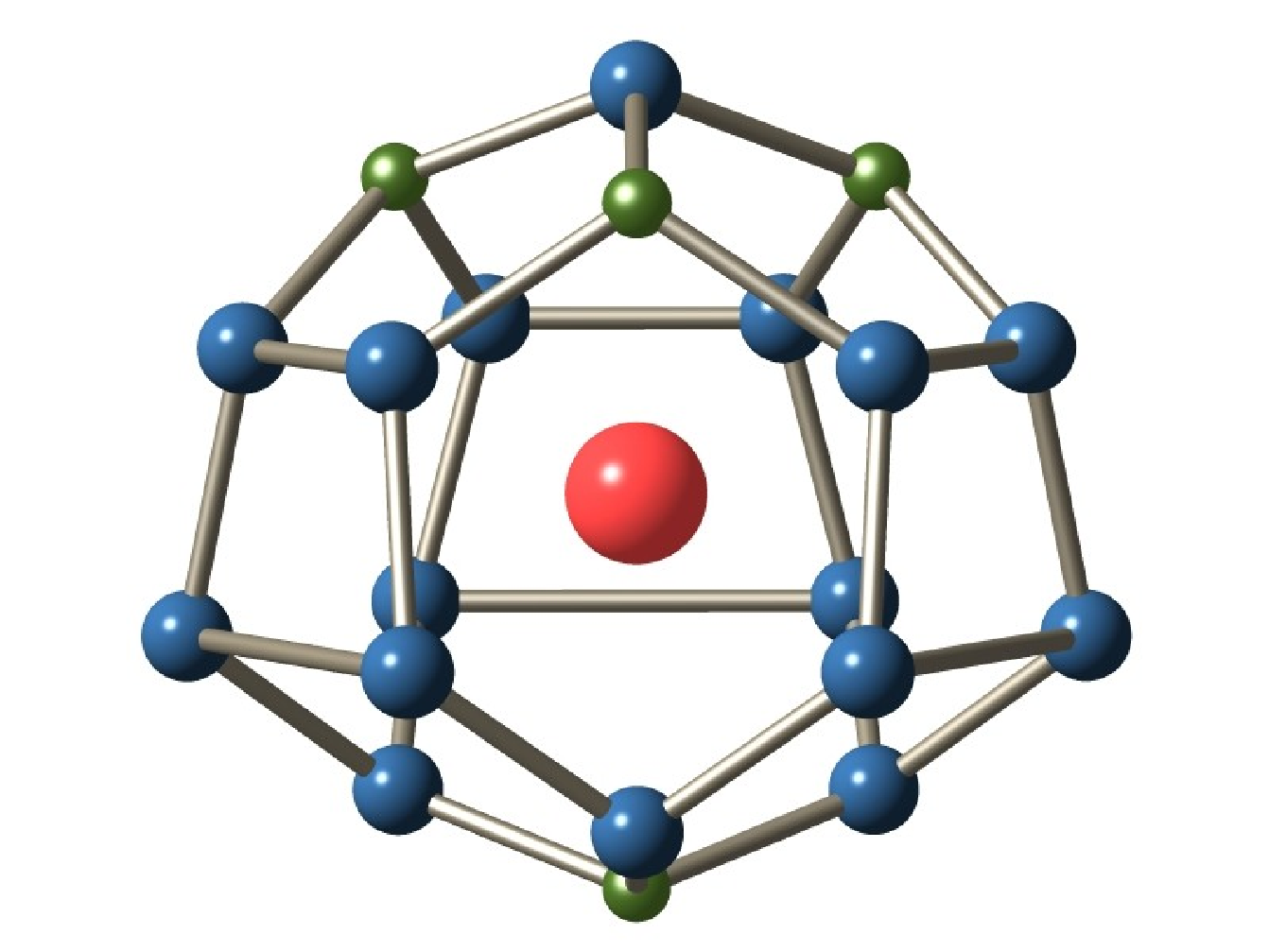}% Here is how to import EPS art
\caption{\label{T8Cage} (Color online) Irregular cage of the
type-VIII clathrate structure adopted by Ba$_8$Ga$_{16}$Sn$_{30}$.
The four smaller cage spheres represent the $8c$ site,
preferentially occupied by Ga atoms.}
\end{figure}

In this work we address the issue by performing single crystal
x-ray diffraction (SCXRD), thermal conductivity $\kappa(T)$ and
heat capacity $C_p(T)$ experiments on Ba$_8$Ga$_{16}$Sn$_{30}$
(BGS) and Ba$_8$Ga$_{16}$Ge$_{30}$ (BGG) crystals, which have been
tuned through the crystal growth process to display n-type or
p-type majority charge carriers as a result of small imbalances in
their Ga:Ge or Ga:Sn ratios.\cite{avi06a,avi06b} By analyzing the
differences and similarities between the behaviors of these
samples, we can test the applicability of the various models
proposed to explain the origin of unusual glasslike behavior, in
this case observed for p-type samples whereas the n-type samples
show the normal crystalline peak.

\section{Experimental Details}

Growth details of the batches used in this work are described in
previous papers.\cite{huo05a,avi06a,avi06b} Single crystalline
polyhedrons of 3-10 mm in diameter were obtained by a self-flux
method. The carrier type is tuned by choosing Ga or Sn
flux\cite{avi06a} in the case of BGS or by adjusting the relative
Ge content in the initial mix with Ga flux\cite{avi06b} in the
case of BGG. The batch name, flux composition and crystal
composition determined by a JEOL JXA-8200 electron-probe
microanalyzer (EPMA) are summarized in table~\ref{EPMA}. The
composition values are averages over 10 regions of each crystal,
although there are random fluctuations of up to $\pm0.1$
throughout the crystals. The values shown for the p-BGG sample
should be considered a correction to those published in refs.
\onlinecite{avi06b} and \onlinecite{umeo05a}, since this was a
more careful evaluation made on the same batch. As expected from
charge balance principles, Ga-rich samples show p-type carriers
while Ge-rich samples show n-type carriers.

\begin{table}
\caption{\label{EPMA} Average Ba:Ga:X content (X = Ge, Sn) in the
four measured crystals as determined by electron-probe
microanalysis.}
\begin{ruledtabular}
\begin{tabular}{ccc}
Sample & Starting Flux & Crystal\\
Name & Composition & Composition\\
\hline
n-BGS & 8 : 16 : 60 & 8.0 : 15.98 : 30.02\\
n-BGG & 8 : 38 : 34 & 8.0 : 15.94 : 30.06\\
p-BGS & 8 : 38 : 30 & 8.0 : 16.14 : 29.86\\
p-BGG & 8 : 38 : 30 & 8.0 : 16.10 : 29.90\\
\end{tabular}
\end{ruledtabular}
\end{table}

Thermal conductivity experiments were performed using a
steady-state method on home-made systems, in the range of
0.3-300~K (BGG) and 4-300~K (BGS), although reliable data is only
obtainable up to about 150 K. At higher temperatures, thermal
losses by radiation and wire conduction prevent the correct
measurement of the intrinsic sample properties. The electronic
contribution $\kappa_c(T)$ of all samples estimated by the
Wiedeman-Franz law is negligible up to 100~K, so the measured
$\kappa(T)$ is equated to the lattice contribution $\kappa_L(T)$.
Heat capacity was measured using a Quantum Design PPMS with its
standard thermal-relaxation method, in the range $0.4\le
T\le300$~K.

For SCXRD experiments, broken pieces of n-BGS and p-BGS with
approximate dimensions of $0.1\times0.1\times0.1$~mm$^3$ were
selected. The diffraction data were collected with a Rigaku R-AXIS
RAPID imaging plate area detector using graphite-monochromated Mo
$K_{\alpha}$ radiation. Refinements were performed using the
CrystalStructure\cite{SCXRD} software. Structures were solved by
direct methods and expanded using Fourier techniques. All sites
were assumed to be fully occupied.

\begin{figure}[tb]
\includegraphics[angle=0,width=86mm]{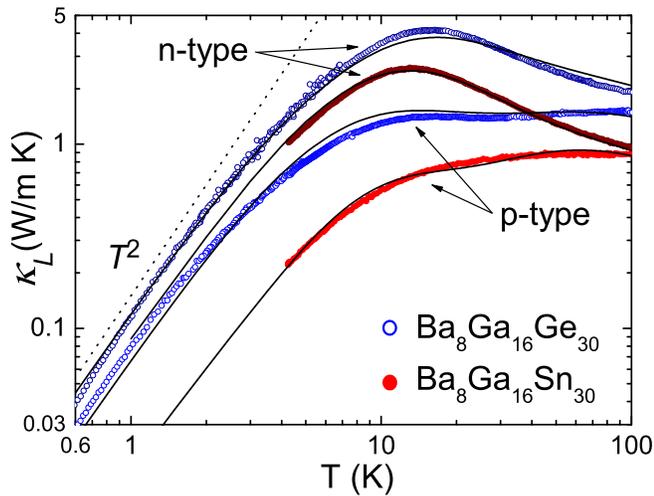}% Here is how to import EPS art
\caption{\label{kappa} (Color online) Temperature dependence of
the thermal conductivity of Ba$_8$Ga$_{16}$Ge$_{30}$ and
Ba$_8$Ga$_{16}$Sn$_{30}$ with different carrier types. Solid lines
are best fits of the TRR model as described in the Discussion
section.}
\end{figure}

\section{Experimental Results}

\subsection{Thermal Conductivity}

The lattice thermal conductivity $\kappa_L(T)$ of all four samples
is shown in Fig.~\ref{kappa} (symbols are the as-measured
experimental data points and solid lines represent fits to the
data by the TRR model which will be detailed in the Discussion
section). At 100 K and above (not shown), $\kappa_L(T)$ for BGS is
roughly half that of BGG (about 1 and 2 W/m K respectively). This
can be understood as a consequence of three main factors: (i) if
the rattling of the guest ions is the main contributor to the
unusually high phonon scattering level in these
materials,\cite{dong01a} the larger cage size in BGS leads to
larger rattling of the guest ions; (ii) in the BGS unit cell all 8
guest ions vibrate with equal intensity (single crystallographic
site for Ba in the Type-VIII clathrate structure), while in BGG
only the 6 guest ions inside the $X_{24}$ cages show large
rattling; and (iii) the heavier Sn atoms produce lower frequency
phonons, which are more easily scattered.

Below 100 K, each sample behaves quite differently, depending more
on the carrier type than on the compound. The two n-type samples
increase towards a peak, while the p-type samples remain at a
plateau, smaller by a factor of $3-4$ in value than the n-type
counterparts near the peak. Below 10 K, $\kappa_L(T)$ for all
samples decreases fast and, in the case of BGG which was measured
to lower temperatures, a gradual crossover to a $T^2$ regime is
clearly observed. This implies a phonon mean free path inversely
proportional to frequency\cite{cohn99a} which is the expected
dependence when phonon scattering by tunnelling states is
dominant. The $T^{2}$ behavior contrasts with a previously
reported result showing a $T^{1.5}$ dependence for p-type
BGG.\cite{ben04a}

\begin{figure}[tb]
\includegraphics[angle=0,width=86mm]{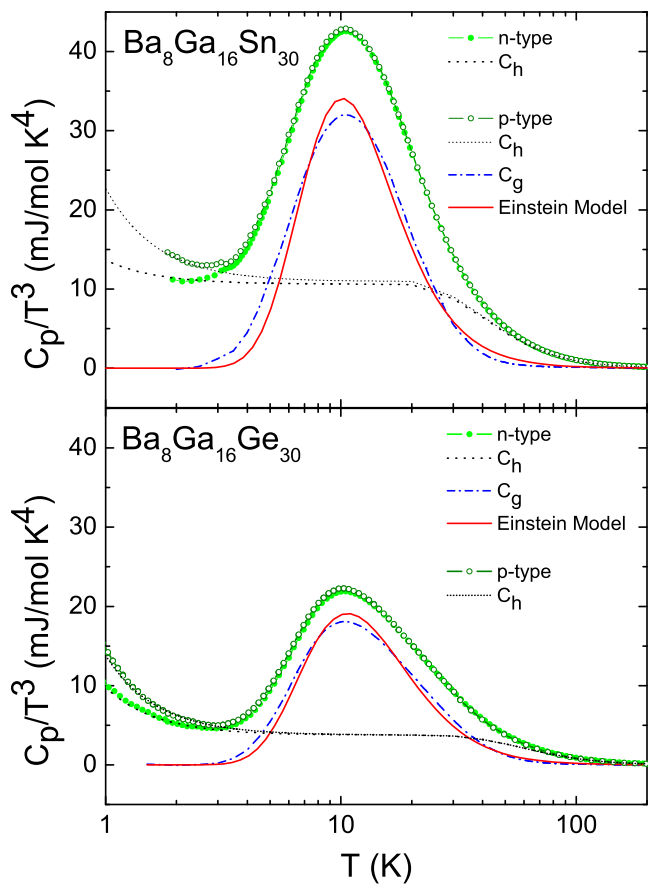}% Here is how to import EPS art
\caption{\label{CpBGX} (Color online) Heat capacity $C_p$ of
Ba$_8$Ga$_{16}$Sn$_{30}$ and Ba$_8$Ga$_{16}$Sn$_{30}$ with
different carrier types, presented as $Cp/T^3 vs. T$.  Solid lines
are best fits of the Einstein model as described in the Discussion
section.}
\end{figure}

\begin{table*}
\caption{\label{n-SCXRD} Summary of crystallographic parameters
from the structural refinement of a n-type
Ba$_8$Ga$_{16}$Sn$_{30}$ single crystal. Space group
$I$\={4}3$n$~(No. 217),~$a=11.586(1)$~\AA ,~$Z=1$, ~$R=0.009$,
~$R_w=0.009$, ~$B_{eq}=(8\pi^2/3) \sum_{i} \sum_{j} U_{ij} a_i^*
a_j^*
 a_i a_j$.}
\begin{ruledtabular}
\begin{tabular}{lllllll}
Atom & site & $x$ & $y$ & $z$ & $B_{eq}$ (\AA$^2$) & occupancy\\
\hline

Ba(1) & 8c & 0.68490(5) & 0.31510(5) & 0.31510(5) & 3.32(1) & 1\\

Ga(1) / Sn(1) & 12d & 0.5000 & 0.0000 & 0.2500 & 1.65(1) & 0.184(12) / 0.816(12)\\

Ga(2) / Sn(2) & 2a & 0.5000 & 0.5000 & 0.5000 & 0.97(2) & 0.158(22) / 0.842(22)\\

Ga(3) / Sn(3) & 24g & 0.41549(4) & 0.14887(4) & 0.41549(4) & 1.333(9) & 0.314(8) / 0.686(8)\\

Ga(4) / Sn(4) & 8c & 0.36558(6) & 0.36558(6) & 0.36558(6) & 1.07(2) & 0.766(12) / 0.234(12)\\
\end{tabular}
\end{ruledtabular}
\end{table*}

\begin{table*}
\caption{\label{p-SCXRD} Summary of crystallographic parameters
from the structural refinement of a p-type
Ba$_8$Ga$_{16}$Sn$_{30}$ single crystal. Space group
$I$\={4}3$n$~(No. 217),~$a=11.587(1)$~\AA ,~$Z=1$, ~$R=0.0218$,
~$R_w=0.0157$, ~$B_{eq}=(8\pi^2/3) \sum_{i} \sum_{j} U_{ij} a_i^*
a_j^*
 a_i a_j$.}
\begin{ruledtabular}
\begin{tabular}{lllllll}
Atom & site & $x$ & $y$ & $z$ & $B_{eq}$ (\AA$^2$) &  occupancy\\
\hline

Ba(1) & 8c & 0.68507(8) & 0.31493(8) & 0.31493(8) & 3.25(2) & 1\\

Ga(1) / Sn(1) & 12d & 0.5000 & 0.0000 & 0.2500 & 1.69(2) & 0.152(16) / 0.848(16)\\

Ga(2) / Sn(2) & 2a & 0.5000 & 0.5000 & 0.5000 & 0.90(3) & 0.233(26) / 0.767(26)\\

Ga(3) / Sn(3) & 24g & 0.41565(4) & 0.14836(4) & 0.41565(4) & 1.308(14) & 0.318(10) / 0.682(10)\\

Ga(4) / Sn(4) & 8c & 0.36577(6) & 0.36577(6) & 0.36577(6) & 1.077(14) & 0.707(12) / 0.293(12)\\

\end{tabular}
\end{ruledtabular}
\end{table*}

\subsection{Heat Capacity}

The data points in Figs.~\ref{CpBGX}a and \ref{CpBGX}b show the
as-measured specific heats $C_p(T)$ for the BGS and BGG samples
respectively, plotted as $Cp/T^3$~$vs.$~$T$. This plotting style
emphasizes the contributions of localized vibrations of guest
atoms (Einstein oscillators), which appear as pronounced peaks
over a ``background'' contribution of a Debye solid. For these
samples, the charge carrier contribution is negligible above
$\sim4$~K, but responsible for the $T^{-2}$ upward curvature upon
cooling below this temperature. A more traditional plot of
$Cp/T$~$vs.$~$T^2$ below 4~K (not shown) is used to estimate with
good accuracy the Sommerfeld coefficient $\gamma$ of the charge
carriers and the Debye temperature $\Theta_D$ of the 46 framework
atoms, and then subtract the host contribution $C_h$ (dotted
lines) in order to isolate the Einstein-like contribution $C_g$ of
the guest ions (dash-dotted peak).

Contrary to the heat transport data, in both BGS and BGG the heat
capacity data show the same behavior above $\sim4$~K for different
carrier types. This result demonstrates there is no fundamental
change in the entropic properties of these compounds within the
range of deviations from stoichiometry studied. If the rattling
behavior of the guest ion is not significantly changed for
different carrier types in the framework, then it should be the
coupling between the guest vibration and the frameworks with
different carriers that changes. In other words, frameworks with
holes have their phonon modes more effectively scattered by the Ba
vibration than those with electrons.

\subsection{Single-Crystal X-Ray Diffraction}

Tables \ref{n-SCXRD} and \ref{p-SCXRD} summarize the refinement
results made for room temperature SCXRD data of n-BGS and p-BGS
respectively. The anomalously large isotropic thermal parameter
$B_{eq}$ of the Ba site compared to the Ga/Sn sites is a signature
of the enhanced vibration (rattling) of the guest ion in the
oversized cage, however, no relevant difference is observed
between the crystals.

The resulting sets of data do not allow a detailed composition
analysis for comparison with EPMA results, because the R factor
was insensitive to occupation probability within deviations of
$\pm0.2$ from stoichiometry, but the framework sites show
consistent preferential occupations for Sn(1) and Sn(2) in the
respective $12d$ and $2a$ crystallographic sites, while Ga(4) has
the preferential occupation of the $8c$ site (in accordance with
the idea that the smaller Ga atom should more easily occupy the
site with smaller bond distances between neighbors) and the $24g$
site remains more randomly occupied by Sn(3) and Ga(3). This is
true for crystals with both types of carriers, the only consistent
and relevant difference we could find in these refinements was a
larger relative occupation of the $2a$ site by Ga(2) for the p-BGS
samples (the top atom in Fig.~\ref{T8Cage}). This could be where
the ``extra'' Ga ions prefer to enter in Ga-rich samples, but
whether or not this can have any influence on the overall
guest/framework coupling would require more detailed
investigation.

\section{Discussion}

We now present and discuss the models used to analyze the data in
Figs. \ref{kappa} and \ref{CpBGX}. The specific heat is expressed
as a sum of 3 main contributions:

\begin{equation}
 C_p = C_c+C_D+C_E%
\label{eq:Cp},
\end{equation}

where $C_c=\gamma T$ is the electronic specific heat of the charge
carriers,

\begin{equation}
 C_D = \frac{12\pi^4N_Dk_B}{5} \int_0^{\Theta_D/T}\frac{x^4e^xdx}{(e^x-1)^2}%
\label{eq:CD}
\end{equation}

with $x=\hbar ck/k_BT$ is the Debye model for the lattice specific
heat of $N_D$ Debye oscillators per unit cell, whose numerical
solution can be found in Solid State Physics textbooks, and

\begin{equation}
 C_{Ei} = \sum_{i}p_iN_{Ei}R\left(\frac{\Theta_{Ei}}{T}\right)^2\frac{e^{\Theta_{Ei}/T}}{\left(e^{\Theta_{Ei}/T}-1\right)^2}%
\label{eq:CE}
\end{equation}

is the Einstein specific heat of the $i$-th vibrational mode of
any existing rattling ions. For our analysis we assume that the 8
rattling guest ions are sufficiently decoupled from the 46 rigid
framework atoms, so that we can make the association $C_g =
C_{Ei}$ and $C_h = C_c + C_D$ respectively.

The solid line in Fig.~\ref{CpBGX}a indicates the best fit of
eq.~\ref{eq:CE} to the isolated Einstein contribution in p-BGS. It
is important to emphasize that in this analysis for the type-VIII
structure, the dimensionality and the number of Einstein
oscillators are fixed at $p=3$ and $N_E=8$ so there is a
\emph{single} fitting parameter $\Theta_E$ for BGS, which alone
governs all the peak characteristics - position, height and width.
The fact that the best fitted curve with $\Theta_E=49.9$~K so
closely reproduces the experimental peak in all these aspects is a
solid testimony to how successful the Einstein model is in
describing the vibrational behavior of the 8 Ba ions in this
compound. The data is actually slightly broadened with respect to
the model, which may be the result of a narrow distribution of
$\Theta_E$ around the mean value (a consequence of Ga/Sn site
disorder), and/or a slightly anisotropic vibration of each guest
ion in its respective irregular cage (resembling an ovoid with
diameter varying between $7.3-8.2$~\AA, see Fig.~\ref{T8Cage}). We
will see next how anisotropic vibration plays a much more
important role in the BGG compound.

Contrary to BGS, the specific heat of BGG \emph{cannot} be
adequately fit with a single $\Theta_{E}$ and $N_E=8$. If $N_E$ is
freed as a fitting parameter, the best fit naturally decreases
this to $N_E=6.1$ (and $\Theta_{E}=55$~K), consistent with the
fact that the 6 Ba(2) ions in the larger $X_{24}$ cages are the
main rattlers, but the fitting quality is still not satisfactory.
Until now, the usual approach\cite{ben04a,umeo05a} to analyze the
heat capacity of BGG has been to assign two different Einstein
contributions ($i=2$ in Eq.~\ref{eq:CE}). This results in
excellent fits with $\Theta_{E1}=70-80$~K and
$\Theta_{E2}=30-40$~K. However, the \emph{number} of Einstein
oscillators $N_{E1}$ and $N_{E2}$ results opposite to what one
would expect if these numbers were to represent the two Ba sites,
i.e., there is a greater number of Ba oscillators with larger
$\Theta_{E1}$ ($N_{E1}=6-9$) than those with smaller $\Theta_{E2}$
($N_{E2}=1.5-2.0$). This is difficult to physically justify, since
larger rattling implies \emph{smaller} $\Theta_{E}$.

We offer an analysis which better reconciles with the guest ions'
known physical properties. The starting point is that, due to the
$X_{24}$ cage shape (Fig.~\ref{T1Cages}, left), the Ba(2) ions
show a strongly \emph{anisotropic} vibration with greater
amplitude within the plane parallel to the cage's two
hexagons.\cite{nol00a} Because the dimensionality $p$ plays a role
in the Einstein model (see Eq.~\ref{eq:CE}), at least two
vibrational modes should be required to describe the Ba(2) site
alone: in-plane ($\Theta_{E2}^\parallel$) and out-of-plane
($\Theta_{E2}^\perp$) respectively. In addition, a third
vibrational mode ($\Theta_{E1}$) is required to account for the
smaller, but still Einstein-like, rattling of the two Ba(1) site
ions in the $X_{20}$ dodecahedra (Fig.~\ref{T1Cages}, right),
which can be assumed isotropic.\cite{sal01a} In this model, the
dimensionalities and numbers of oscillators are predefined:
$p_1N_{E1}=3\times2$, $p_2^\parallel N_{E2}^\parallel=2\times6$,
$p_2^\perp N_{E2}^\perp=1\times6$, so the fitting parameters are
only the three Einstein temperatures, with the additional
constraint that
$\Theta_{E2}^\parallel<\Theta_{E2}^\perp,\Theta_{E1}$. The best
fitting of this model results in $\Theta_{E1}=87.2$~K,
$\Theta_{E2}^\parallel=49.4$~K and $\Theta_{E2}^\perp=87.1$~K. The
similarity between $\Theta_{E2}^\parallel$ of BGG and $\Theta_{E}$
of BGS is reasonable, since the largest diameters of both cages
are essentially the same ($\sim8.2$~\AA). The similarity between
$\Theta_{E2}^\perp$ and $\Theta_{E1}$ in BGG is also reasonable
since the $X_{24}$ cage size in the out-of-plane direction is very
close to the $X_{20}$ cage size ($\sim5.5$~\AA). This means a
further simplification can be made in the model by assuming only
two parameters $\Theta_{E1}$ and $\Theta_{E2}$ with
$p_1N_{E1}=p_2N_{E2}=12$, where $\Theta_{E1}$ represents the 3D
vibration of the Ba(1) ions and the 1D out-of-plane vibration of
the Ba(2) ions; while $\Theta_{E2}$ represents the larger, 2D
in-plane vibration of the Ba(2) ions. This results in the solid
curve shown in Fig.~\ref{CpBGX}b and, as with BGS, the data for
BGG is only slightly broadened with respect to the model.

With the heat capacity parameters determined, we now analyze the
lattice thermal conductivity $\kappa_L$ of all samples, using the
same procedure applied previously for analysis of the n-BGS
sample,\cite{huo05a} which is in turn based on the TRR model
initially used in ref.~\onlinecite{cohn99a} to describe
Sr$_8$Ga$_{16}$Ge$_{30}$. In the semi-classical theory, $\kappa_L$
is given by

\begin{equation}
 \kappa_L = \frac{1}{3}\int_0^{\omega_D}d\omega[C_L(\omega,T)vl]%
\label{eq:kappa},
\end{equation}

where $C_L(\omega,T)$ is the phonon specific heat, $\omega_D$ is
the Debye frequency, $v$ is the average sound velocity and $l$ is
the phonon mean free path, which must be averaged over all major
contributing scattering mechanisms. Thus, in the TRR model it is
written as

\begin{equation}
 l = (l_{TS}^{-1}+l_{Res}^{-1}+l_{Ray}^{-1})^{-1}+ l_{min}%
\label{eq:mfp}.
\end{equation}

The low-energy excitations of the guest ions tunnelling between
localized states scatter phonons as

\begin{equation}
 l_{TS}^{-1} = A\left(\frac{\hbar\omega}{k_B}\right)\text{tanh}\left(\frac{\hbar\omega}{2k_BT}\right)+\frac{A}{2}\left(\frac{k_B}{\hbar\omega}+\frac{1}{B T^3}\right)^{-1}%
\label{eq:lTS},
\end{equation}

where A and B are microscopic parameters describing the tunnelling
states characteristics.\cite{grae86a} At higher energies, phonons
are scattered through a resonance effect against guest ion
rattling as:

\begin{equation}
 l_{Res}^{-1} = \sum_{i}\frac{C_i\omega^2T^2}{\left(\omega_i^2-\omega^2\right)^2+\Gamma_i\omega_i^2\omega^2}%
\label{eq:lRes},
\end{equation}

where $C_i$ and $\Gamma_i$ are phenomenological parameters related
to a simple mechanical oscillator.\cite{pohl62a} We also need to
include the empirical but always present, frequency-only dependent
Rayleigh scattering

\begin{equation}
 l_{Ray}^{-1} = D\left(\frac{\hbar\omega^4}{k_B}\right)%
\label{eq:lRay},
\end{equation}

and finally the last term $l_{min}=1$~\AA $ $ is the cut-off
limit.

Results from the best fits of the data shown in Figs. \ref{kappa}
and \ref{CpBGX} are summarized in Table~\ref{FITS}. The most
relevant results in terms of comparing the p-type with n-type
samples are the increase in the resonant scattering level ($C_i$),
and in the TS scattering level. The latter can be expressed by the
ratio $A/B=\tilde{n}(\hbar v)^2/\pi k_B$, which in glasses is
essentially a measure of the \emph{subset} density of tunnelling
states $\tilde{n}$ that are able to strongly couple to the phonons
and effectively scatter them.\cite{grae86a} Therefore, the
increase in A/B observed upon changing from n-type to p-type cages
does not necessarily mean the total density of TS has increased,
only that the existing states are more effectively coupled.

\begin{table}
\caption{\label{FITS} Parameters used to generate the solid line
curves in Figs. \ref{kappa} and \ref{CpBGX}, which best fit the
respective experimental data set for lattice thermal conductivity
and specific heat. See text for detailed descriptions.}
\begin{ruledtabular}
\begin{tabular}{cccccc}
Symbol & Unit & n-BGG & p-BGG & n-BGS & p-BGS\\
\hline
$A$ & $10^4$/(m K) & 1.4 & 2.5 & 2.5 & 17\\
$B$ & 1/K$^2$ & 0.1 &  0.1 & 0.1 & 0.1\\
$A/B$ & $10^5$K/m & 1.4 & 2.5 & 2.5 & 17\\
$C_1$ & 1/(m s$^2$ K$^2$) & 0.2 &  2.0 & 0.7 & 5.0\\
$\Theta_{E1}$ & K & 87 &  87 & 50 & 50\\
$\Gamma_1$ &  & 0.5 &  1.5 & 0.4 & 1.5\\
$C_2$ & 1/(m s$^2$ K$^2$) & 0.2 &  2.0 & - & -\\
$\Theta_{E2}$ & K & 49\footnote{Two-dimensional vibration (see text).} & 49\footnotemark[1] & - & -\\
$\Gamma_2$ &  & 0.5 &  1.5 & - & -\\
$D$ & K$^{4}$/m & 0.85 &  0.5 & 2.8 & 1.7\\
$\gamma$ & mJ/(mol K$^2$) & 6 &  9 & 1.3 & 11\\
$\Theta_D$ & K & 288 &  288 & 200 & 200\\
$v$ & m/s & 2898 &  2898 & 2250 & 2250\\
\end{tabular}
\end{ruledtabular}
\end{table}

An interesting exercise can be made to help understand the
influence of these different contributions in the TRR model. If we
begin with the fitting results for $\kappa_L(T)$ of the n-type
samples, it is impossible to fit the respective p-type
$\kappa_L(T)$ by increasing the intensity of only one of these
contributions (TS or resonant). The TS are mainly responsible for
decreasing the low-temperature $\kappa_L(T)$ up to the first few
Kelvins, and by itself the TS contribution is incapable of
changing the peak into a plateau. Conversely, an increase in the
resonant scattering level (based on \emph{fixed} values of
$\Theta_{E}=50$~K from heat capacity and increased
phenomenological coupling strength parameters $C_i$), readily
brings the peak down to a plateau/dip, but quickly loses its
ability to follow the $\kappa_L(T)$ drop below about 10~K.
Therefore, we may conclude that the TRR model adequately
reproduces the entire range of $\kappa_L(T)$ for all samples up to
100~K, provided that the coupling of the framework phonons with
the guest ion tunnelling \emph{and} rattling is increased in
p-type samples.

Let us now focus on some other proposals regarding candidate
mechanisms for glasslike behavior in clathrates, which challenge
the TRR model. First: is the presence of tunnelling states really
necessary, or is \emph{off-center} vibration a sufficient
mechanism? The question was raised by Bridges and
Downward\cite{bri04a} primarily based on the existing data at the
time, where Sr and Eu guests clearly show off-center
sites\cite{cha00a,sal01a,zer04a} and glasslike $\kappa(T)$, while
Ba guests appear to show \emph{on-center}
vibration\cite{cha00a,sal01a,zer04a} (within experimental
resolution) and a crystalline peak. Later studies demonstrated
glasslike behavior for p-type BGG\cite{ben04a,avi06b} and now for
p-type BGS (this work), so this argument by itself is no longer
valid, unless a closer look at the Ba vibration in these compounds
through microscopic techniques shows that off-center vibration
modes do exist for p-type samples (even if much smaller than for
Sr and Eu guests) but not for n-type samples. Raman scattering and
EXAFS studies are presently being conducted on our carrier-tuned
BGG crystals, which may help clarify this issue. Still, good
arguments were made by the authors in terms of describing how
off-center vibration does indeed help enhance the \emph{coupling}
between guest vibration modes and the framework
phonons.\cite{bri04a,baum05a}

A second challenge to the TRR model is: can the shift from
crystalline to glasslike behavior be explained solely by phonon
scattering mechanisms within the framework, i.e., by interactions
between phonons and charge carriers? This question was raised in a
series of papers by Bentien \emph{et
al.}\cite{ben04a,ben05a,ben06a} which we now discuss.

The first work\cite{ben04a} called attention to an observed
$\kappa(T)\propto T^{1.5}$ dependence at low-temperature for p-BGG
and a kink in their data at about 2~K (neither of which were
reproduced with our crystals). They also pointed out that the
phonon-charge carrier mechanism could not explain the lowering of
$\kappa_L(T)$ above $\sim15$~K, so the resonant scattering on the
guest vibration was once again invoked, but to account for only
the differences above 15~K. The second work\cite{ben05a} compared
several polycrystalline samples of type-VIII and type-I
Eu$_8$Ga$_{16}$Ge$_{30}$ ($\alpha$-EGG and $\beta$-EGG
respectively,\cite{pas01a} all with n-type carriers), clearly
demonstrating that $\beta$-EGG shows glasslike $\kappa_L(T)$ while
$\alpha$-EGG does not. The difference was interpreted in terms of
changes in the band structure, with a much enhanced effective mass
$m^*$ found in $\beta$-EGG. However, the cage sizes and shapes are
also quite different between these two structures. The type-I
$X_{24}$ cages are essentially the same size for all Ge clathrates
($5.5\times8.2$~\AA, see Fig.~\ref{T1Cages}) but the type-VIII
cage in $\alpha$-EGG ($6.7\times7.5$~\AA $ $ ovoid similar to
Fig.~\ref{T8Cage}) is significantly smaller than that of BGS, so
any change in $\kappa_L(T)$ can also be argued or modelled in
terms of changes in the Eu vibration modes and their coupling to
the framework. Unfortunately $\alpha$-EGG samples with p-type
carriers are as yet unavailable, but it wouldn't be surprising if
they showed glasslike $\kappa_L(T)$ as we found in p-BGS. The
third and more recent work\cite{ben06a} shows results for
Ba$_8$Ni$_x$Ge$_{46-x}$ similar to what we have obtained here for
BGG and BGS, therefore the same analysis and discussion we have
conducted here can also be applied to those results.

Still, it is obvious that the influence of charge-carriers cannot
be neglected with respect to their density $n_c$, effective mass
$m^*$, electronic mean free path $l_c$, etc. It is quite clear
from our measurements and all previously reported data on
Ba-filled clathrates, that the p-type carriers are playing a
relevant role in producing an increased phonon scattering in these
compounds, which we view as yet another additional factor capable
of contributing to lower $\kappa_L(T)$, possibly through direct
interaction with the phonons, but especially by mediating an
enhanced coupling of these with the guest vibration modes. A few
brief examples for such mediation possibilities are: 1) n-type
frameworks could allow a greater degree of \emph{coherence} in the
vibrations of neighboring Ba guests than p-type frameworks, which
would lead to larger mean free paths and less effective
scattering; 2) Since the type and density of charge carriers
result from stoichiometry imbalances, they may affect the
framework rigidity at certain sites, and therefore how easily it
can couple with the rattler ions.

\section{Conclusion}

We have succeeded in growing large single crystals of
Ba$_8$Ga$_{16}$Sn$_{30}$ and Ba$_8$Ga$_{16}$Ge$_{30}$ with both
n-type and p-type majority carriers, and found that these
compounds show low temperature lattice thermal conductivity
behavior strongly dependent on the carrier type. A shift from
crystalline to glasslike behavior is observed for both compounds
when changing the majority carriers from n-type to p-type through
composition tuning. These differences can be mostly reproduced by
an increase in resonant scattering, however, an increase in both
resonant and tunnelling scattering levels are required to
reproduce the full set of data below 100 K. Heat capacity and
single-crystal x-ray diffraction data indicated that these
increases are not the result of any major change in the guest
ions' vibrational behaviors, therefore a more effective coupling
of the frameworks with p-type carriers to the TS and rattling
vibrations of the guest ions is the most likely mechanism. The
$T^2$ dependence in $\kappa_L(T)$ obtained at lowest temperatures
for both n-type and p-type Ba$_8$Ga$_{16}$Ge$_{30}$ indicates that
tunnelling states should be present for the Ba(2) ions in this
compound, therefore its mere presence is insufficient to guarantee
glasslike $\kappa_L(T)$.

In fact, our results indicate that the various proposed mechanisms
which may lead to glasslike behavior are all partially correct and
at the same time incomplete. The general scenario that we see
emerging can indeed be expressed as: \emph{it's all about the
coupling}. For reasons that still need to be explained
microscopically, the n-type frameworks are more weakly coupled to
the guest vibration modes than the p-type frameworks. Thus, the Ba
ions' smaller and (almost?) on-center vibration is not coupled
strongly enough to the n-type framework phonons to produce the
glasslike behavior, but the p-type framework crosses the necessary
coupling strength threshold to achieve this scattering regime. In
contrast, Sr and Eu ions in the type-I Ge clathrates have clearly
off-center and larger rattling, capable of a strong enough
coupling even with the n-type frameworks to produce glass-like
behavior (no p-type frameworks have been reported yet for these
compounds). In a series of carefully tuned Ba-based clathrates it
should be possible to observe a continuous transition from
glasslike to crystalline $\kappa_L(T)$. Likewise, in a series of
n-type (Sr,Eu)-based clathrates the same continuous transition
should be observed not from carrier tuning, but from a physical or
chemical reduction of cage size to dampen the off-center vibration
level.

\begin{acknowledgments}
We thank M. Udagawa, F. Bridges and D. Huo for fruitful
discussions, and Y. Shibata for the EPMA analysis. Specific heat
and thermal conductivity measurements were carried out at Natural
Science Center for Basic Research and Development (N-BARD)
Hiroshima University. This work was financially supported by the
Grants in Aid for Scientific Research(A) (No.~18204032), the COE
Research (13CE2002) and the priority area ``Skutterudite''
(No.~15072205) from MEXT, Japan.
\end{acknowledgments}

%\bibliography{condmat,clathrates}% Produces the bibliography via BibTeX.

\end{document}